\documentclass[10pt,journal,compsoc]{IEEEtran}

\usepackage{color}

\usepackage{array}
\usepackage{multirow}
\usepackage{todonotes}

\ifCLASSOPTIONcompsoc
  \usepackage[nocompress]{cite}
\else
  \usepackage{cite}
\fi
\ifCLASSOPTIONcompsoc
    \usepackage[caption=false, font=normalsize, labelfont=sf, textfont=sf]{subfig}
\else
    \usepackage[caption=false, font=footnotesize]{subfig}
\fi

\ifCLASSINFOpdf
  \graphicspath{{../pdf/}{../jpeg/}}
  \DeclareGraphicsExtensions{.pdf,.jpeg,.png}
\else
  \usepackage[dvips]{graphicx}
  \graphicspath{{../eps/}}
  \DeclareGraphicsExtensions{.eps}
\fi
\newenvironment{commentwrapper}[1]{\color{#1}}{\color{black}}

\usepackage{xcolor}
\definecolor{RED}{rgb}{0.7,0.0,0.0}

\begin{document}
%
\title{WristSketcher: Creating Dynamic Sketches in AR with a Sensing Wristband }
%
%
%
%

\author{Enting Ying,
        Tianyang Xiong,
        Shihui Guo,
        Ming Qiu,
        Yipeng Qin,
        Hongbo Fu%
}
\IEEEtitleabstractindextext{%
\begin{abstract}

Restricted by the limited interaction area of native AR glasses (e.g., touch bars), it is challenging to create sketches in AR glasses. Recent works have attempted to use mobile devices (e.g., tablets) or mid-air bare-hand gestures to expand the interactive spaces and can work as the 2D/3D sketching input interfaces for AR glasses.
Between them, mobile devices allow for accurate sketching but are often heavy to carry, while sketching with bare hands is zero-burden but can be inaccurate due to arm instability.
In addition, mid-air bare-hand sketching can easily lead to social misunderstandings and its prolonged use can cause arm fatigue.
As a new attempt, in this work, we present WristSketcher, a new AR system based on a flexible sensing wristband for creating 2D dynamic sketches, featuring an almost zero-burden authoring model for accurate and comfortable sketch creation in real-world scenarios.
Specifically, we have streamlined the interaction space from the mid-air to the surface of a lightweight sensing wristband, and implemented AR sketching and associated interaction commands by developing a gesture recognition method based on the sensing pressure points on the wristband.
The set of interactive gestures used by our WristSketcher is determined by a heuristic study on user preferences.
Moreover, we endow our WristSketcher with the ability of animation creation, allowing it to create dynamic and expressive sketches.
Experimental results demonstrate that our WristSketcher i) faithfully recognizes users' gesture interactions with a high accuracy of 96.0\%; ii) achieves higher sketching accuracy than {\it Freehand} sketching; iii) achieves high user satisfaction in ease of use, usability and functionality; and iv) shows innovation potentials in art creation, memory aids, and entertainment applications.
\end{abstract}

}

\maketitle

\IEEEdisplaynontitleabstractindextext

%
\IEEEpeerreviewmaketitle

\ifCLASSOPTIONcompsoc
\IEEEraisesectionheading{\section{Introduction}\label{sec:introduction}}
\else
\section{INTRODUCTION}
\label{sec:introduction}
\fi

%
%
%
%
\begin{figure*}[h!]
    \centering
    \includegraphics[width=\linewidth]{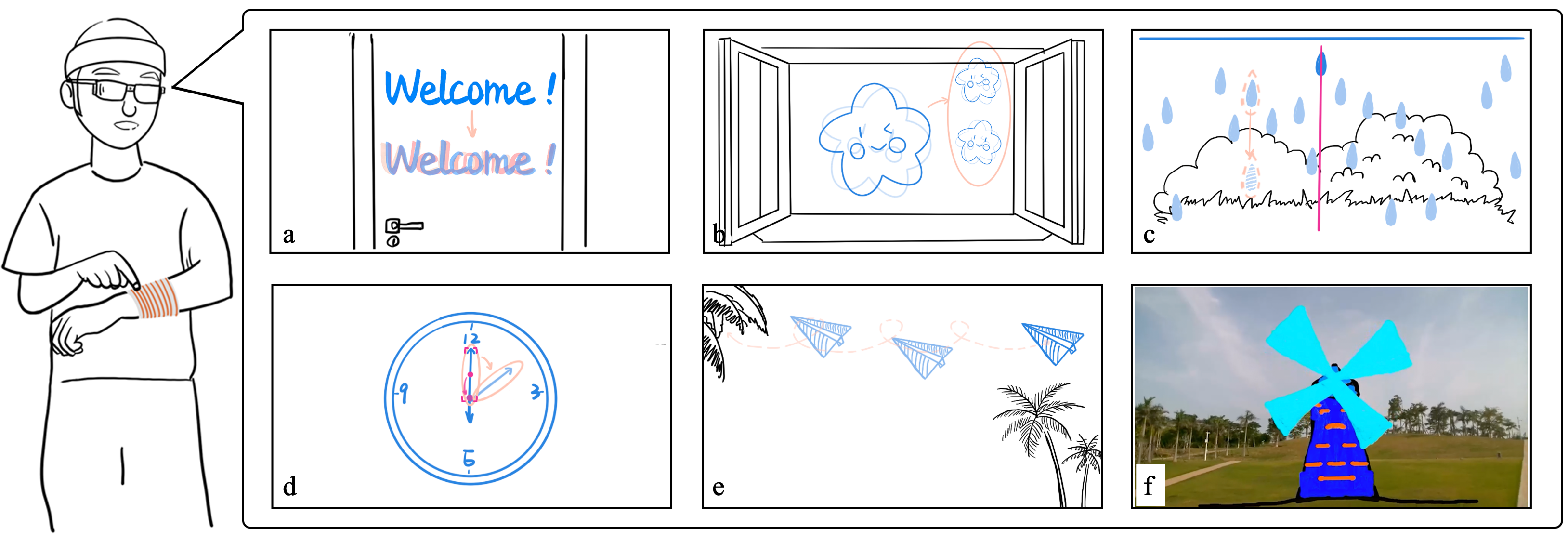}
    \caption{WristSketcher supports dynamic sketch creation for AR glasses by using a flexible pressure sensing wristband. 
    Our WristSketcher allows users to sketch in real-world scenes, and add five types of animation effects: (a) \emph{Doodle}, (b) \emph{Frame}, (c) \emph{Emit}, (d) \emph{Rotate}, and (e) \emph{Move} to the sketches. For example, (f) is a windmill sketched using our WristSketcher with a rotation animation added to its blades.} 
    \label{fig:main}
\end{figure*}
\IEEEPARstart{S}{ketching} is a natural language for all individuals to express their thoughts, feelings, and emotions. 
When used in an Augmented Reality (AR) environment, sketching is provided with a context of surrounding scenes. Compared to traditional creation methods, AR-based creation inspires users' creativity and can thus be used as narrative tools for storytelling \cite{Pronto}, teaching aids \cite{RealitySketch, kang2020armath}, 3D design tools \cite{SymbiosisSketch,Mobi3DSketch}, etc.
Such a wide range of applications show the infinite expressive power of sketching in AR.

Early works on AR sketching are mostly based on smart phones or tablets, which are effective but not immersive
\cite{Mobi3DSketch}, \cite{A:JustaLine}, \cite{A:ARDraw}, \cite{A:ARXPaint}, \cite{A:ARDrawAugmentedReality3D}. 
In contrast, directly sketching with native AR glasses is immersive but can be extremely troublesome due to the limited interaction area of their touch bars.
To address this issue, existing works resort to either gesture recognition and tracking techniques that allow for bare-hand sketching in the air \cite{ARSketch,Bare-Handed2018}, or mobile devices such as tablets \cite{WhatYouSketchIsWhatYouGet,SymbiosisSketch} as input agents.
Between them, bare-hand sketching in 3D space is zero-burden but lacks physical support, which has been shown to be inaccurate \cite{ExperimentalEvaluation}, causing arm fatigue and social misunderstandings \cite{PocketThumb}. In contrast, mobile devices allow for ergonomic and accurate sketching, but are heavy, limiting their application scenarios.

In this paper, we draw ideas from flexible electronics and propose WristSketcher, a novel system (Fig.~\ref{fig:main}) based on a flexible sensing wristband for creating 2D dynamic sketches in AR.
Our WristSketcher comprises a lightweight flexible film on a user's wrist to detect the user's touch pressure in real time, leading to an almost zero-burden authoring model for accurate and comfortable sketching in real-world scenarios.
We also design a set of sketching associated interaction commands based on a heuristic study and implement them with a pressure-based gesture recognition method.
In addition, inspired by RealitySketch~\cite{RealitySketch}, we endow our WristSketcher with the ability of animation creation, allowing the creation of dynamic and more expressive sketches.
Integrating the above, we implement a dynamic sketching application with diverse functions such as \emph{Recognition}, \emph{Draw}, \emph{Erase}, and \emph{Animation}.
Finally, we justify the effectiveness of our WristSketcher through i) its high gesture recognition accuracy; ii) its superiority in accuracy compared to {\it Freehand} sketching, that is one of the modes of mid-air bare-hand interaction, the input is made by tracking the position of the hand in real time; iii) a positive user evaluation which demonstrates that our WristSketcher provides a pleasant and comfortable creative experience, and iv) its innovation potentials in art creation, memory aids, and entertainment applications.

In summary, our main contributions include:
\begin{itemize}
  \item [1.]
  We present WristSketcher, a novel 2D sketch input agent for AR based on a flexible sensing wristband, which is lightweight and portable, giving users a pleasant and comfortable creative experience.
  \item [2.]
  We design a set of interaction commands for our WristSketcher based on a heuristic study. They are intuitive, easy-to-remember, accurate, and efficient. We implement them with a pressure-based gesture recognition method.
  \item [3.]
  We demonstrate the superiority of our WristSketcher with i) a comparison with \emph{Freehand} sketching on sketching accuracy; ii) a user study, which shows that our WristSketcher can accurately recognize ($\sim$96\% accuracy) and respond to users' interaction by touching on its surface. In addition, we have received positive feedback from users in terms of ease of use, usability, and functionality of our WristSketcher. These feedback also show the innovation potential of our WristSketcher in art creation, memory aids, and entertainment applications.
\end{itemize}

\section{RELATED WORK}
\subsection{Sketching Input in AR/VR}

\textbf{Bare-Hand.} Thanks to the advance in deep learning, gesture recognition, and tracking are now well-established technologies and have been widely deployed in real-world applications. 
For example, instead of introducing third-party proxy tools, many AR/VR applications in modern mixed reality headsets \cite{GestureKnitter} adopt a bare-hand interaction method, which is natural and expressive and can support sketching~\cite{AiRSculpt, HoloARt, Bare-Handed2018, ARSketch}. 
Despite its efficiency and immersive user experience, bare-hand sketching in AR/VR can be inaccurate (148\% less than 2D sketch input) due to the lack of physical support~\cite{ExperimentalEvaluation, Multiplanes}.
Furthermore, it can lead to social misunderstandings when used in the public and its prolonged use can cause arm fatigue.

\textbf{Pen (in-the-air).} 
Pens are popular input devices for both AR and VR \cite{isthepenmightier}. 
Some works \cite{CavePainting,VRSketchPen,FreeDrawer} show the high expressiveness of pens in sketching using AR/VR headsets. 
To facilitate accurate drawing, the sizes, shapes, and weights of these pens are usually carefully designed \cite{VRSketchPen,isthepenmightier}.
However, since they are used in-the-air, pens also suffer from social misunderstandings and arm fatigue.

\textbf{Mobile Devices (Positioning).} AR Development Kits like ARKit and ARCore have empowered mobile devices (e.g., mobile phones, tablets) with spatial awareness \cite{ARKitandARCore}. 
Based on them, Just-a-Line \cite{A:JustaLine}, ARDraw \cite{A:ARDraw} and AR Draw Augmented Reality 3D \cite{A:ARDrawAugmentedReality3D} all make use of the global position of a mobile phone for 3D sketching.
In this way, the mobile devices are used both for display and for sketching (positioning), and are thus not immersive.
When used for positioning in the air, mobile devices suffer from social misunderstandings and arm fatigue as well.

\textbf{Mobile Device (Touching).} 
To obtain an immersive experience, mobile devices, especially tablets, can be used together with AR/VR headsets where the headsets are used for display and the mobile devices are only used for sketch input. For example, \cite{VRSketchIn,SymbiosisSketch,WhatYouSketchIsWhatYouGet} make use of the touch screens of the mobile devices for sketch input. Thanks to the physical support of the screen, these methods allow for more accurate sketching than above methods. 
However, these devices are heavy, making it inconvenient to carry around.

\textbf{Wearable Devices.}
Wearable devices are promising solutions for various real-world problems including AR sketching.
For example,
Qian et al.\cite{Portalware} utilize a wearable device (camera) that can be worn on the fingertip to extend the field of view for mobile AR sketching.
Jiang et al.\cite{HandPainter} use Manus VR gloves and extend the hand point positions to include fingertips to improve the accuracy of VR sketching. However, their gloves are only used as a physical proxy and their system still relies on optical perception and positioning.

In this work, we present WristSketcher, a novel 2D sketching system for AR based on a flexible sensing wristband. 
Compared to in-air interactive input in AR, we provide users with a supportable interactive surface. Users can interact with the sketching application in the AR glasses through gestures on the surface of our WristSketcher to complete AR sketches. This surface-based creation method will support users in creating more accurate AR sketches. In addition, compared to using other agents as input, our WristSketcher as a wearable device can be worn on the surface of a body, which provides convenience for 
going out to a certain extent. For example, users do not need to hold it in their hands when not creating, and the lightweight device feature can also give users a comfortable experience.

\subsection{Dynamic Sketching}
Compared to static ones, dynamic sketches are those with animated effects and are thus more expressive. How to add such animations efficiently has been an important topic in human-computer interaction (HCI).
Recently, Suzuki et al. \cite{RealitySketch} define the two modes of dynamic sketching, \emph{Separated} and \emph{Embedded}, according to their design spaces, i.e., whether the sketches are embedded in real-world scenes.
With this definition, \cite{kitty,Mixed-Initiative,Draco,LiveSketch} follow the \emph{Separated} mode and isolate dynamic sketching from the real world. 
\cite{Pronto,PoseTween} proposed a dynamic sketching method for videos, by adding sketches to an input video and controlling their motion according to keyframes. 
Note that their methods still adopt the \emph{Separated} mode as they rely on keyframes of pre-recorded videos rather than embedding sketches into real-world scenes in real-time. 
DoodleLens \cite{A:DoodleLens} follows the \emph{Embedded} mode, by converting sketched animation frames drawn on paper into digital content, placing them in real-world scenes, and animating them by quickly switching frames. 
RealitySketch \cite{RealitySketch} goes a step further in the \emph{Embedded} mode, allowing drawn lines and defined prefabs to interact with real-world events in real-time.

In this work, we follow the \emph{Embedded} mode and endow our WristSketcher with the ability to create dynamic sketches with AR glasses. In addition, we carefully design its associated interaction commands to enable the creation of custom animations with rich options in a user-friendly way.

\subsection{Interaction with Flexible Wearable Devices}

Flexible wearable devices typically consist of textiles, sensing units, and the data transmission and processing systems.
Like clothing, such devices can be worn on various parts of a human body, such as the head \cite{UseYourHead}, hands \cite{Glove}, wrists \cite{Skill-Sleeves}, thighs \cite{PocketThumb}, feet \cite{Sensock}, leading to various applications in motion analysis \cite{VirtualLocomotion,BodyMotion}, interactive control \cite{I/OBraid}, healthcare \cite{ProCover}, etc.

In previous works, flexible wearable devices complete the input of instructions by providing diverse input semantics. 
For example, Dobbelstein et al.\cite{PocketThumb} and Heller et al. \cite{FabriTouch} implement fabric tactile input by embedding capacitive multi-touch sensing devices into fabrics and using the front pocket of the pants as a touch area. 
Dobblelstein \cite{Belt} propose a touch belt for Google Glass input, whose large input surface allows users to interact in a subtle and unobtrusive way without raising their arms.
GestureSleeve \cite{gesturesleeve} is a wristband variant that supports the recognition of tap and swipe gestures. 
zPatch \cite{ZPatch} is an eTextile patch that enables hover, touch and pressure input via resistive and capacitive sensing, and is used to control music players, text input and game input.
Despite these works, the use of wearable devices in AR sketching and their associated interaction commands is still an under-explored topic. 

In this work, we try the application scene of dynamic sketch creation by using a sensing wristband in AR.

\section{SYSTEM OVERVIEW}

\subsection{Prototype Setup}
\begin{figure} [!h]
    \centering
    \includegraphics[width=\linewidth]{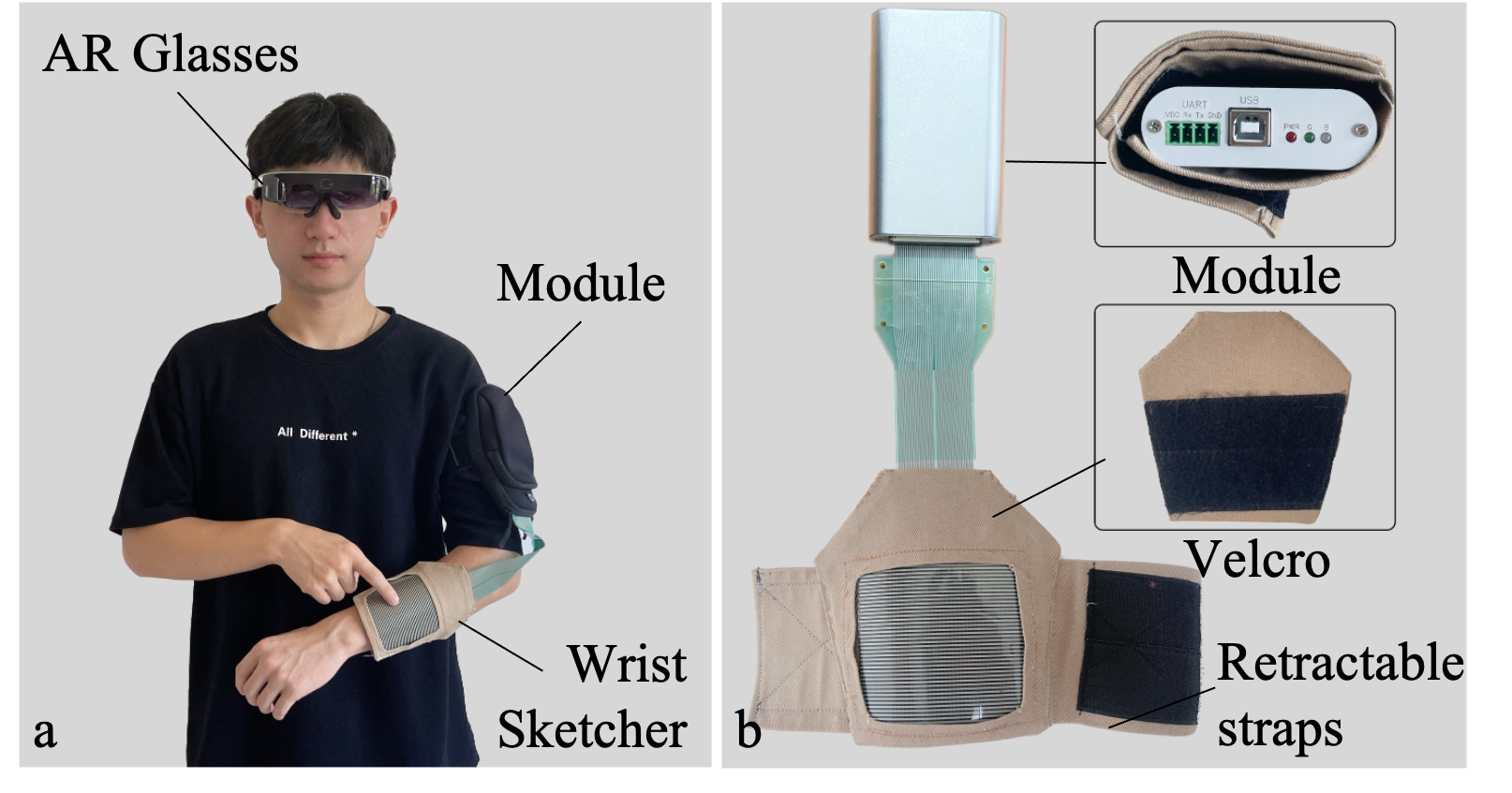}
	  \caption{(a) AR glasses and our WristSketcher worn on a wrist. (b) WristSketcher's detailed structure. Module: signal acquisition module.}
	  \label{fig:protype} 
\end{figure}
\begin{figure*}[!htbp]
 \centering
 \includegraphics[width=\linewidth]{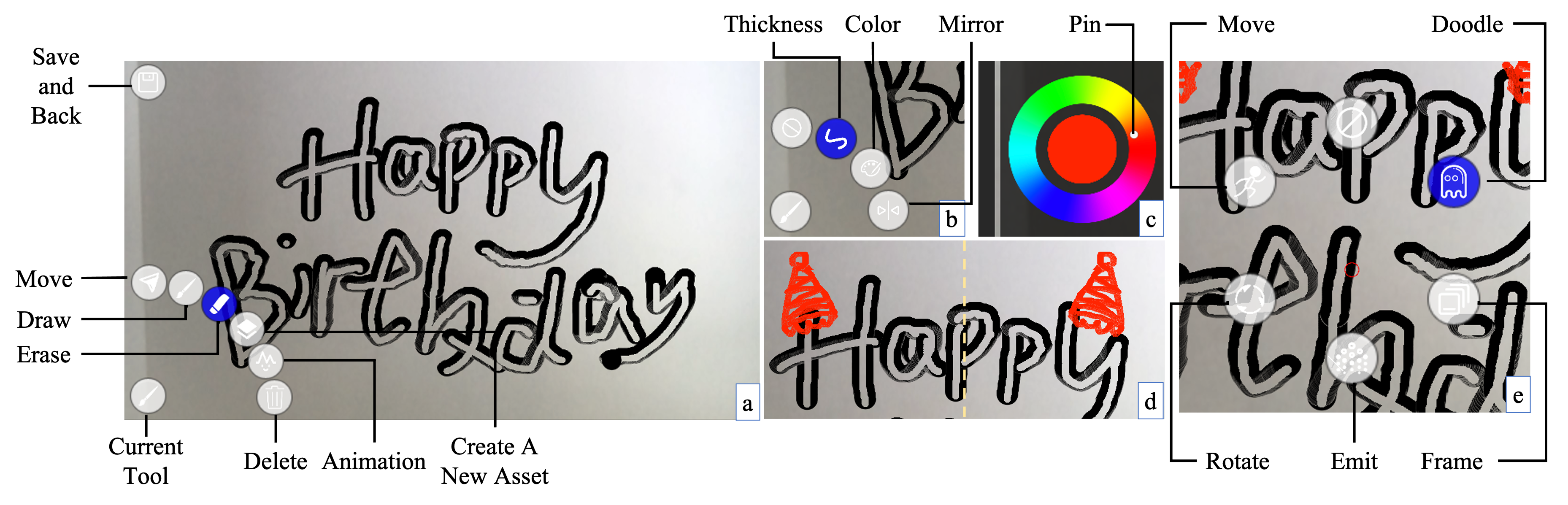}
 \caption{
 The user interface of our WristSketcher. (a) 
 The main menu with functions: \emph{Move}, \emph{Draw}, \emph{Erase}, \emph{Create a New Asset}, \emph{Animation}, and \emph{Delete}.
 (b) The sub-menu of \emph{Draw} with functions: \emph{Change Thickness}, \emph{Change Color}, and \emph{Mirror Brush}. 
 (c) The color picker. 
 (d) Illustration of the \emph{Mirror Brush}, which simultaneously draws two hats that are symmetric to the midline of the canvas. 
 (e) The animation panel consisting of five animation effects: \emph{Doodle}, \emph{Frame}, \emph{Emit}, \emph{Rotate}, and \emph{Move}.}
 \label{fig:interface}
\end{figure*}

\textbf{WristSketcher.} As illustrated in Fig. \ref{fig:protype}a, the proposed WristSketcher is composed of a flexible pressure film and a signal acquisition module (``Module'' for short). The former is a sensing point array sensor made of polyester film with excellent comprehensive mechanical properties, highly conductive materials, and nano-scale varistor materials, with a size of 8.36 × 8.36$\mathrm{cm}^{2}$ and a thickness of 0.2$\mathrm{mm}$ [9] (Fig. \ref{fig:protype}b). It is divided into a bottom flexible circuit pressure-sensitive layer and a top flexible circuit pressure-sensitive layer, which are pasted by double-sided tape to isolate the upper and lower sensing areas. When the sensing area is pressurized, the two pressure-sensitive layers are in contact with each other, and the output resistance of the channel varies with the position. The film has a total of 1,936 sensing points, each with a size of 1.6 × 1.6$\mathrm{cm}^{2}$ and arranged with a gap of 0.3$\mathrm{mm}$. The minimum pressure is 20$\mathrm{g}$, and it can respond within 10$\mathrm{us}$. During the experiment, in order to avoid data interruption, we selected 40 sensing points in the horizontal direction and 40 in the vertical direction. The velcro is located on the back of the film (Fig. \ref{fig:protype}b), and can help us attach the film to the surface of the retractable straps tied to the human body, so as to carry it with us. The signal acquisition module is a data collector with a serial interface and USB interface (Fig. \ref{fig:protype}b), which can collect pressure information at a frame rate of 60FPS. In our current implementation, we use a USB interface to connect the collector to a computer equipped with a serial data processing program.

\textbf{AR Glasses.} We use Rokid Glass 2 \cite{A:Rokid} (Fig. \ref{fig:protype}a), an optically transparent head-mounted AR glasses launched by Rokid Corporation Ltd in 2020, to develop our system. This device has a 1280 $\times$ 720 resolution screen, is based on Android system, and supports USB-C connection. 
Due to the lack of a SLAM function in the current device, positioning sketch elements in real-world scenes is challenging. To address this issue, we use a method based on image recognition and tracking, 
which identifies useful contents using (relevant patches cropped from) the original images captured by the camera.

\textbf{Data Processing Unit.} In this work, we use a USB interface to connect the collector to the computer (Intel(R) Core(TM) i7-6500U CPU @ 2.50GHz 2.59 GHz, RAM 8.00 GB) equipped with a serial data processing and a recognition algorithm program. We then use socket communication to transmit the user's interaction information to the AR glasses.

\subsection{Sketching User Interface}
The software of our WristSketcher can be divided into two modules: i) AR recording and rendering, ii) sketching interaction.
We implement them in Unity, and use Vuforia technology to identify and track real-world scenes. 
We introduce the user interface and functions of ii) as follows.

\subsubsection{Sketching Interaction}

The user interface of our WristSketcher software is shown in Fig. \ref{fig:interface}. The menu is hidden by default, except for \emph{Current Tool} and \emph{Save and Back}, to reduce the occlusion caused by the menu bar and thus provide a wider visual interaction area. The main menu (Fig. \ref{fig:interface}a) shows the basic sketch creation tools: \emph{Move}, \emph{Draw}, \emph{Erase}, \emph{Create a New Asset}, \emph{Animation}, and \emph{Delete}. Users activate the menu and select menu items through gestures. Each menu has a corresponding sub-menu (Fig. \ref{fig:interface}b), and the current functional sub-task can be executed by interacting with the sub-menu, as described below. Besides, \emph{Undo} can be executed by a specific gesture so we deliberately remove it from the interface for a more concise view.

\textbf{Color Picker.} In \emph{Draw}, the color picker (Fig. \ref{fig:interface}c) is called up by a shortcut command. When the color picker is displayed, the user can move the pin (Fig. \ref{fig:interface}c) to select a color. 
This color is reserved for the next stroke and works as the indicated color of the brush.
Note that the color picker is hided when sketching, allowing for a larger canvas.

\textbf{Mirror Brush.} This function is used to assist in creating a symmetrical shape (Fig. \ref{fig:interface}d). When this property is checked, strokes drawn by the user on the canvas will be mirrored.

\textbf{Animation Component.} The animation component panel is shown in Fig. \ref{fig:interface}e, which enables users to create five types of animations. By selecting an animation item, the corresponding type of animation will be bound to the current asset. 
Note that users can not only add, but also overlay animation styles to create more engaging and dynamic sketches.
Details of the five types of animations are described below.

\subsubsection{Animation Creation}
To create engaging and dynamic sketches, we draw ideas from BABA Is You\cite{A:BaBaisyou}, Kitty\cite{kitty}, Draco\cite{Draco}, PoseTween\cite{PoseTween}, and introduce five types of animations (\emph{Doodle}, \emph{Frame}, \emph{Emit}, \emph{Rotate}, and \emph{Move}) into our WristSketcher. 
Tab. \ref{table:animationfunctions} shows the properties of each type of animation.
\begin{table}[!ht]
\centering
\setlength{\tabcolsep}{10mm}
\caption{Properties of each type of animation.}
\begin{tabular}{|c| c|} 
 \hline
 Animation & Property \\
 \hline Doodle & -  \\ 
 \hline \multirow{2}{3em}{Frame} & New resource \\& Frame rate \\  
 \hline \multirow{3}{3em}{Emit} & Ejection area \\& Movement trajectory \\
 \hline \multirow{3}{3em}{Rotate} & Rotation angle\\& Rotation time \\& Rotation center\\
 \hline \multirow{2}{3em}{Move} & Movement trajectory \\  &Movement time\\ 
 \hline
\end{tabular}
\label{table:animationfunctions}
\end{table}

\begin{figure}[!h]
 \centering
 \includegraphics[width=\linewidth]{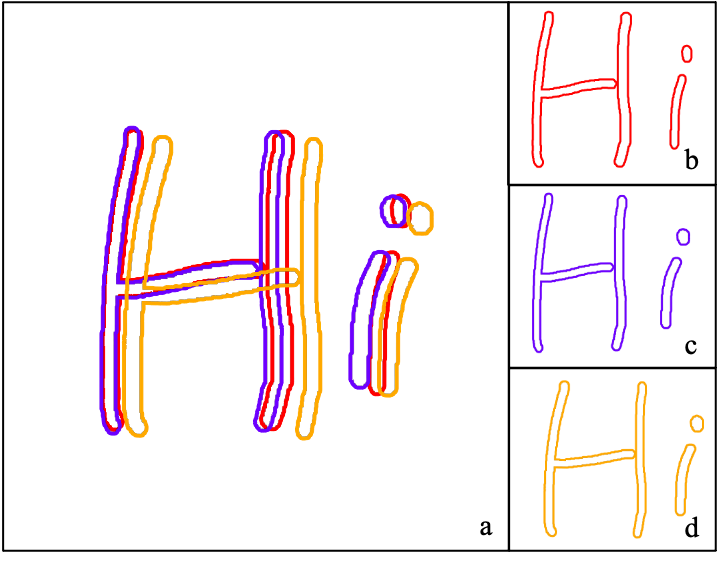}
 \caption{
 Doodle animation (dithering) on ``Hi''. (a) Overlay plot of three consecutive frames. (b) Frame 1. (c) Frame 2. (d) Frame 3.
 }
 \label{fig:doodle}
\end{figure}

\textbf{Doodle.} \emph{Doodle} is a constantly dithering image effect. 
Users can make their sketches jitter (Fig. \ref{fig:doodle}) by simply binding \emph{Doodle} to their input static sketches. 
Whereas achieving such an effect in a traditional animation pipeline may require a large number of hand-drawn frames, our \emph{Doodle} requires only a static sketch, thereby saving massive amounts of time.

\begin{figure}[!h]
 \centering
 \includegraphics[width=\linewidth]{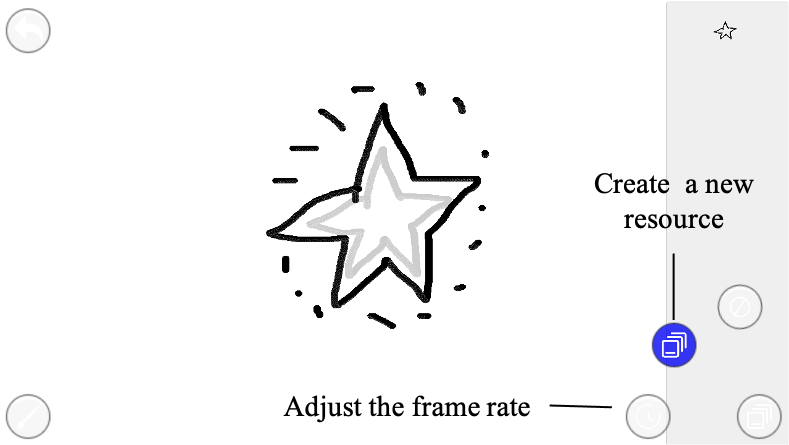}
 \caption{
 Frame animation of twinkling stars. Two stars (i.e., black and gray) are created and switched in turn. Note that the Frame menu supports creating new resources and adjusting the frame rate.
 }
 \label{fig:frame}
\end{figure}

\textbf{Frame.} \emph{Frame} is a frame animation plug-in (Fig. \ref{fig:frame}) to achieve the fast-page turning effect. Such frame-based animations are commonly used in animation scripting and other fields.
When using \emph{Frame}, users can create multiple frames, and animate them by fast switching.
In addition, users can customize the frame rate.

\begin{figure}[!h]
 \centering
 \includegraphics[width=\linewidth]{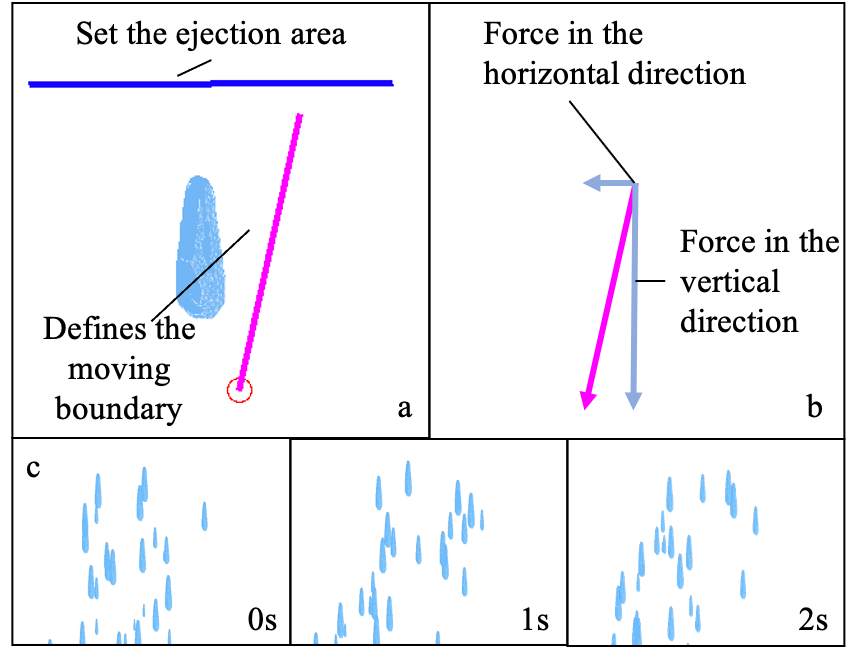}
 \caption{
 (a) Raindrop: the sketched object to be ejected; Blue line: emitter; Red line: the direction guides the object's movement, which can be regarded as ``force'' . (b) Decomposed the force in the horizontal and vertical directions. (c) The final raining effect, with raindrops falling from the emitter in a curved trajectory.
 }
 \label{fig:emit}
\end{figure}

\textbf{Emit.} \emph{Emit} uses the currently created asset as a material and continuously emit it. It has three functions: spray area setting, motion trajectory setting, and gravity simulation. The process of creating an Emit animation is shown in Fig. \ref{fig:emit}a. The user first creates a line (the blue one) to set an ejection area, from which the object will be ejected with an initial velocity, and then defines the movement trajectory of the object by creating a straight line (the red one). For a straight line to guide the motion of an object, we further introduce ``force''. Similar to those in mechanics, the straight line in Fig. \ref{fig:emit}b can be viewed as a resultant force from the the horizontal and vertical forces. 
The final effect is shown in Fig. \ref{fig:emit}c. 
The raindrops follow the trajectory and spray area initially defined by the user to simulate the effect of raindrops being blown by the wind. 

\textbf{Rotate.} 
\emph{Rotate} generates rotation effects by continuously modifying the rotation properties of an object.
By default, we set the rotation point to be the center of mass of an object.
The user can also customize the positions of rotation points, e.g., rotating clock hands around a clock's center.
The angle and time of rotation are controlled by two sliding bars, respectively.

\textbf{Move.}
Binding \emph{Move} to an asset allows it to move along a predetermined trajectory. In \emph{Move}, users can set the trajectory and movement time of the object.

\section{STUDY 1: HEURISTIC STUDY}

The interaction design for our WristSketcher can be challenging due to the lack of visual correspondence between AR display and gesture positioning on the wristband. To obtain a set of easy-to-learn and intuitive gestures to input sketches and interactive commands, we follow the common practice in HCI (e.g., surface computing \cite{User-DefinedGestures}, mobile interaction \cite{GestOnHMD,EarBuddy}, and augmented reality \cite{One-DimensionalHandwriting}), and conduct a heuristic study on users as follows, which can improve the learnability and usability of gestural user interfaces \cite{User-DefinedMotionGestures}.

\subsection{Participants}
We invited 26 participants (16 males, 10 females), aged 18-25 years ($\overline{x}= 22.73$, SD = 1.85), to participate in this experiment, one of whom was left-handed. 
Among them, 18 participants majored in computer science and the remaining were from different disciplines. All the participants frequently used 2D touch interfaces in their daily lives. The participants did not communicate with each other during the experiment, and were paid \$10 each at the end of the experiment.



\subsection{Procedure}
\label{sec:procedure}
During the experiment, we only provided the participants with WristSketcher, as well as the designed user interface. Users can simulate interactive tasks on WristSketcher, but cannot get any timely feedback. Upon arrival, the participants were introduced to the purpose of the experiment by the facilitator and asked to complete a pre-study questionnaire to obtain relevant information about them and then voluntarily sign a consent form.
The facilitator then introduced the device (our WristSketcher and the AR glasses) for this experiment and showed the user interface to the participants. 
Since it is impractical for users to design gestures for all the functions of the system, and some users may not have the ability to design systematically, we have classified the functions that are frequently invoked in the system and formed the following questions: 
\begin{itemize}
    \item Q1: Please design a gesture command to activate the main menu.
    \item Q2: Please design a gesture command to activate the secondary menu.
    \item Q3: Please design a gesture command to activate the tertiary menu.
    \item Q4: Please design a gesture command to switch menu items.
    \item Q5: Please design a gesture command to release the menu.
    \item Q6: Please design a gesture command as a confirmation after the operation.
    \item Q7: Please design a gesture command as an undo operation.
\end{itemize}
The participants were directed to answer these questions and were asked the reasons for their answers. After answering all the questions, they could review and revise their answers to ensure that there were no conflicts between instructions.

\begin{figure}[!h]
 \centering
 \includegraphics[width=\linewidth]{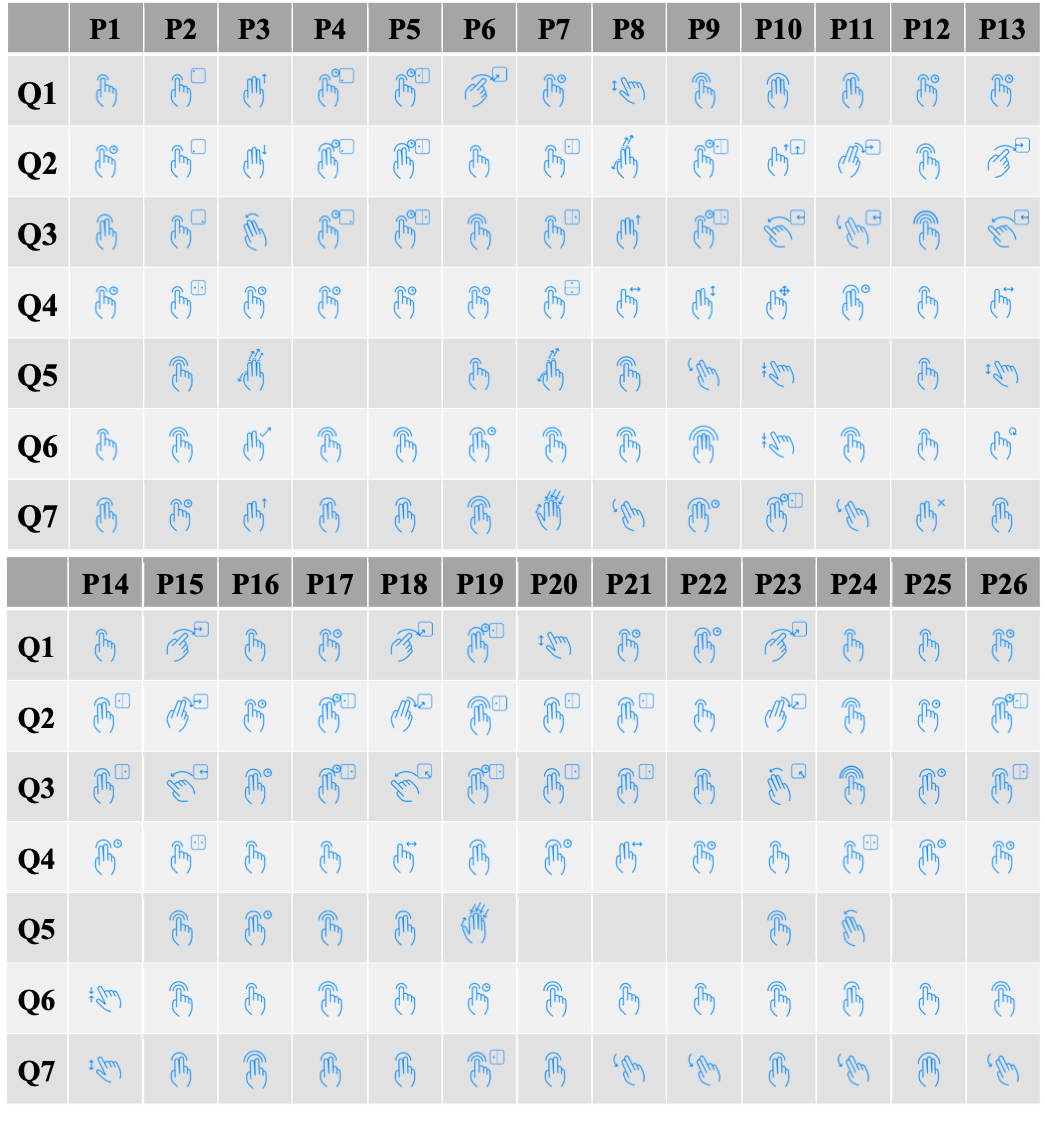}
 \caption{The touch gestures designed by the 26 participants for the 7 questions in Sec.~\ref{sec:procedure}. The point in the rectangle indicates the position of the first touch.}
 \label{fig:heuristic}
\end{figure}

Fig. \ref{fig:heuristic} shows the touch gestures designed by the participants for the 7 questions above. In addition to the gestures commonly used in 2D touch devices, including \emph{Tap}, \emph{Long Press}, and \emph{Slide}, the participants also designed some custom gestures, such as \emph{Five-Finger Pinch} and \emph{Five-Finger Open}. For the various responses from the participants, we found that even when designing the gesture for the same question, different participants have different opinions and preferences. In order to meet the participants' preferences as much as possible, we analyze these results from macro and micro perspectives as depicted below.

\subsubsection{Macro Analysis}
\begin{figure}[!h]
\centering
	  \subfloat[]{
       \includegraphics[height=3.9cm]{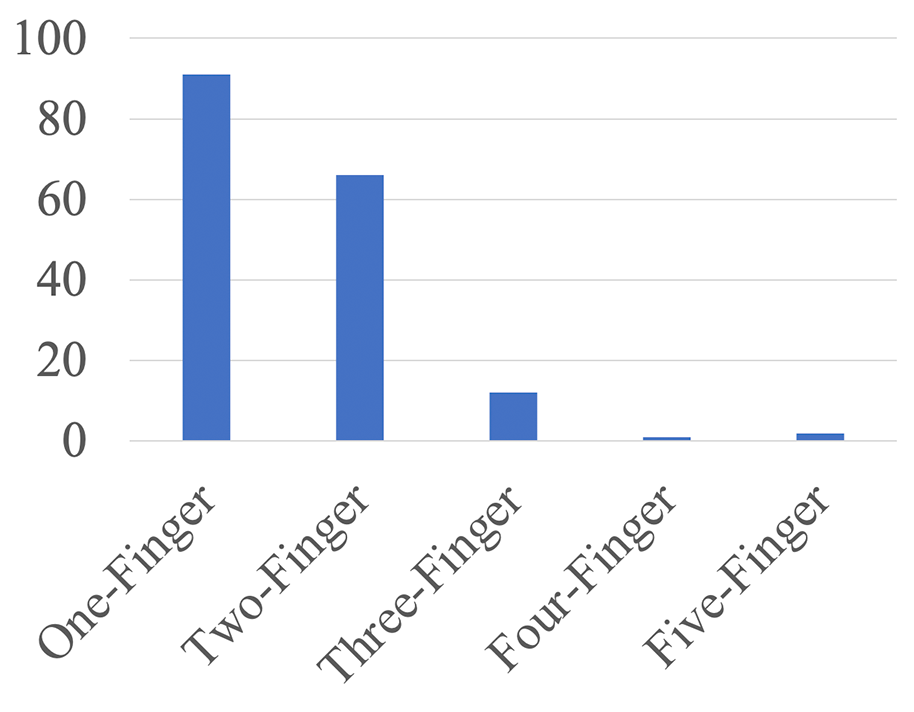}}
    \label{fig:fingercount}
	  \subfloat[]{
        \includegraphics[height=3.9cm]{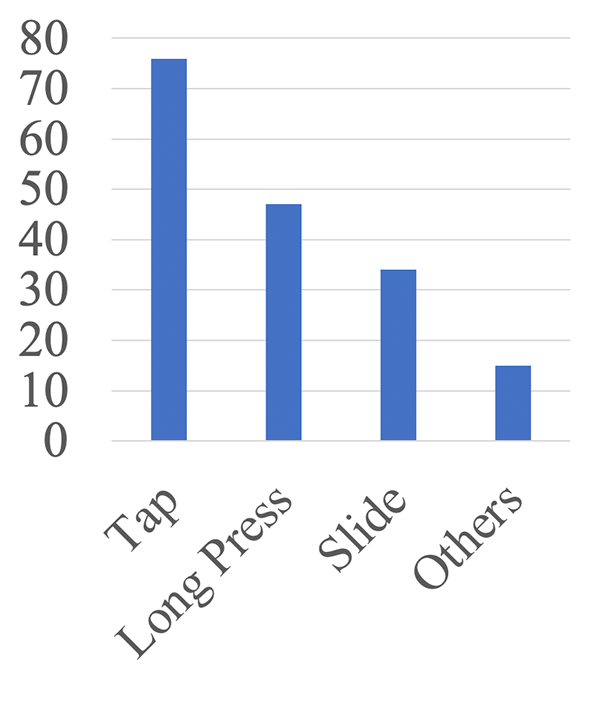}}
    \label{fig:GestureType}
 \caption{(a) The result of classifying the gestures according to the number of fingers involve.(b) The result of classifying the gestures according to the type.}
 \label{fig:macro}
\end{figure}

\textbf{Finger Count.} We classified the user-designed gestures according to the number of fingers involved. 
As shown in Fig. \ref{fig:macro}a, there is a clear preference over using fewer fingers for interaction, which is consistent with the participants' daily habits of using \emph{One-Finger} and \emph{Two-Finger} for touch interaction.
Therefore, we prioritized the use of fewer fingers when designing interaction gestures.

\textbf{Gesture Type.} 
We also classified the participants-designed gestures into four categories according to their types: \emph{Tap}, \emph{Long Press}, \emph{Slide}, and \emph{Others}.
As shown in Fig. \ref{fig:macro}b, among the four interaction gestures, \emph{Tap} (44\%) is the most popular among the participants, followed by \emph{Long Press} (27\%) and \emph{Slide} (20\%). 
Since \emph{Slide} could easily be confused with \emph{Draw}, we dropped it in our final design.

\subsubsection{Micro Analysis}
Our 7 questions can be grouped into four categories according to their relationships, namely Q1-Q3 (menu activation), Q4-Q5 (menu operation), Q6 (end interaction), and Q7 (undo). 
Interestingly, we observed that the participants tended to select the same type of gestures for all activities in the same group. We analyze their details as follows.

\begin{figure}[!h]
\centering
	  \subfloat[]{
       \includegraphics[height=3.9cm]{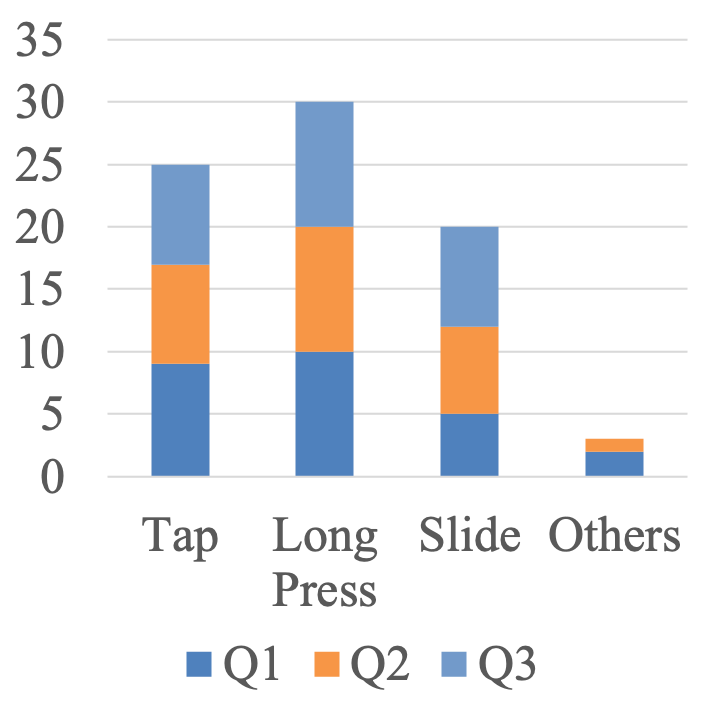}}
    \label{fig:Q1Q3}
	  \subfloat[]{
        \includegraphics[height=3.9cm]{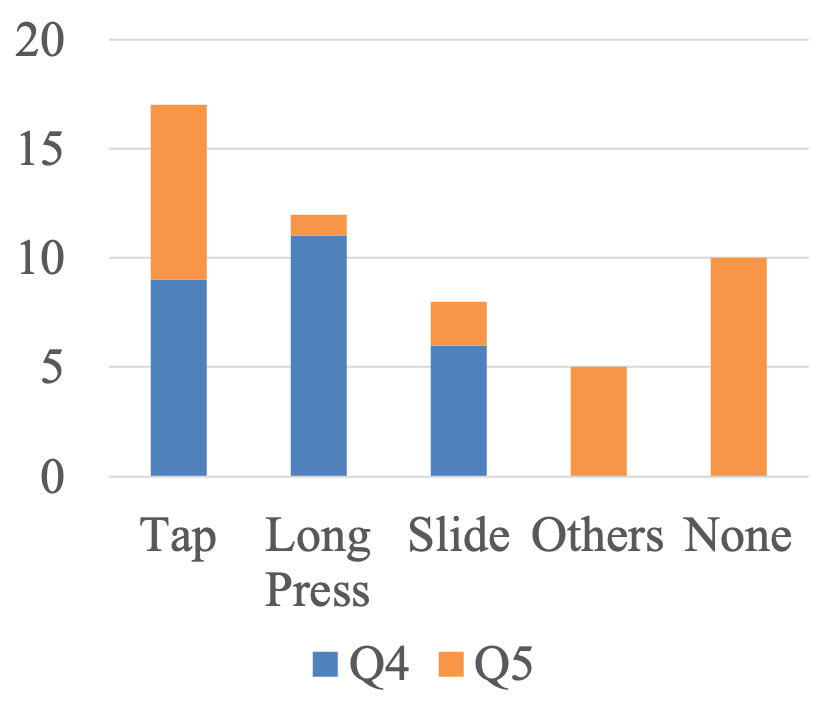}}
    \label{fig:Q4Q5}
    \subfloat[]{
        \includegraphics[height=3.9cm]{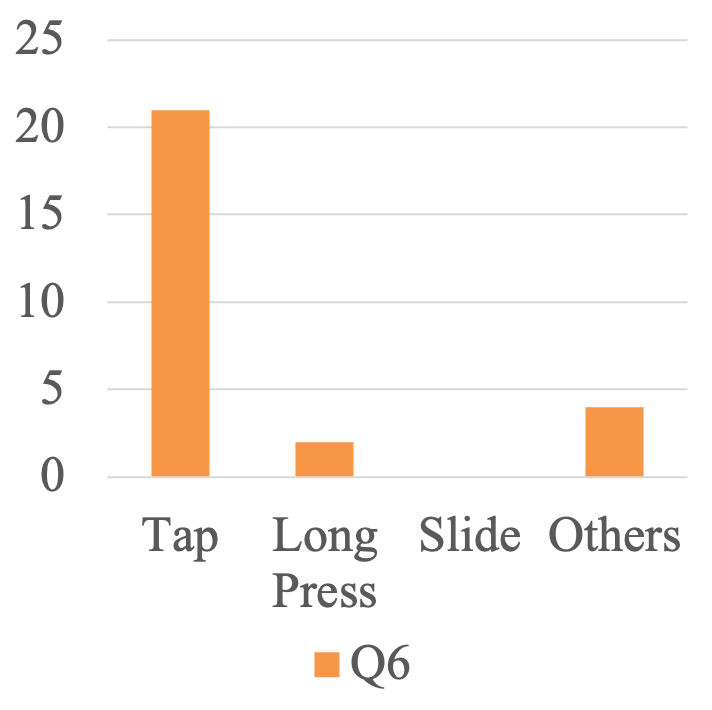}}
    \label{fig:Q56}
    \subfloat[]{
        \includegraphics[height=3.9cm]{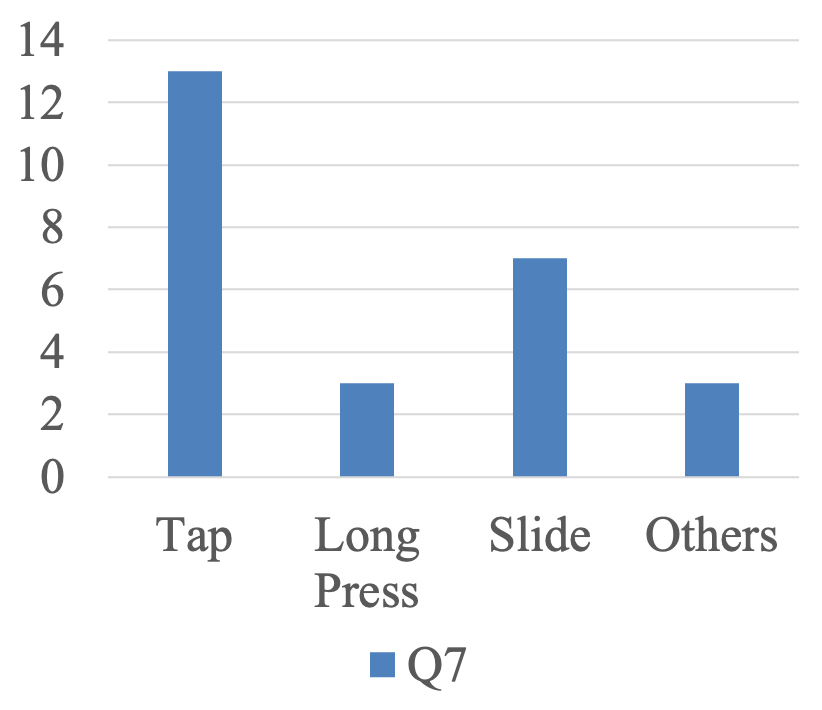}}
    \label{fig:Q7}
 \caption{(a) The results of Q1-Q3. (b) The results of Q4-Q5. (c) The results of Q6. (d) The results of Q7.}
 \label{fig:qifadata}
\end{figure}

\textbf{Q1-Q3.} Q1-Q3 are about how to activate the menu. Fig. \ref{fig:qifadata}a shows the aggregated results of selected types of gestures for these questions.
It can be observed that for menu activation, \emph{Long Press} (38\%) ranks first. Compared with \emph{Tap} (32\%), using \emph{Long Press} has two benefits: i) it reduces the impact of accidental touches; ii) it avoids the confusion between \emph{Tap} and drawing a small dot.

\textbf{Q4-Q5.} Q4-Q5 are about menu operations. 
As shown in Fig. \ref{fig:qifadata}b, when the menu is activated, the largest number of participants preferred using \emph{Long Press} (42\%) to switch menu items. 
Observing similar patterns in Q1-Q3, we believe one major reason for this is a habit inherited from the participants' preferences in menu activation. Interestingly, from their feedback, some participants who choose other options in Q1-Q3 switched to \emph{Long Press} as they felt the automatic switching of menu items reduced the interaction burden. 
For the release of the menu, the participants have a high preference over ``None'', i.e., issuing the releasing command when a persistent state is changed (e.g., \emph{Long Press}). This allows them to remember one less gesture, and we adopt this in our final design.

\textbf{Q6.} Q6 is about ending an interaction. 
As Fig. \ref{fig:qifadata}c shows, the participants show similar preferences over the interaction gestures for Q6: \emph{One-Finger Double Tap}, which is a vivid gesture for operation confirmation.

\textbf{Q7.} Q7 is about undoing an operation. Note that we assign a global gesture to it because \emph{Undo} complements \emph{Erase} and is frequently used in the drawing process. 
As Fig. \ref{fig:qifadata}d shows, the participants preferred using \emph{Tap} to \emph{Undo}. Inspired by the digital illustration software Procreate \cite{A:Procreate}, we use \emph{Two-Finger Tap} for \emph{Undo} in our final design.

The correspondence between the gestures and interaction commands of our final design is shown in Tab. \ref{table:correspondencegeco}.
\begin{table}[htb]
\caption{The correspondence between gestures and interaction commands.}  
\begin{center}
\begin{tabular} { | m{4cm}<{\centering}|m{4cm}<{\centering} |}
\hline   \textbf{Gesture} & \textbf{Command} \\   
\hline  \emph{One-Finger Left Long Press} & Activate the main menu \\
\hline  \emph{Two-Finger Left Long Press} & Activate the secondary menu \\
\hline  \emph{One-Finger Right Long Press} & Activate the tertiary menu\\
\hline  \emph{Hold On} & Switch the menu items \\
\hline  \emph{Cancel Long Press} & Release the menu\\
\hline  \emph{One-Finger Double Tap} & Confirm \\
\hline  \emph{Two-Finger Tap} & Undo \\
\hline  \emph{Two-Finger Right Long Press} & Switch between \emph{Move} with \emph{Draw} \\
\hline  \emph{One-Finger (Left/Right/Top/Bottom) Tap} & The property value successively -1/+1/+1/-1 \\
\hline  \emph{One-Finger (Left/Right/Top/Bottom) Long Press} & The property value is continuously -1/+1/+1/-1\\
\hline
\end{tabular}
\label{table:correspondencegeco} 
\end{center}   
\end{table}

\section{IMPLEMENTATION TECHNIQUES}
\subsection{Touch Point Recognition}
 \begin{figure}[tb]
 \centering 
 \includegraphics[width=\columnwidth]{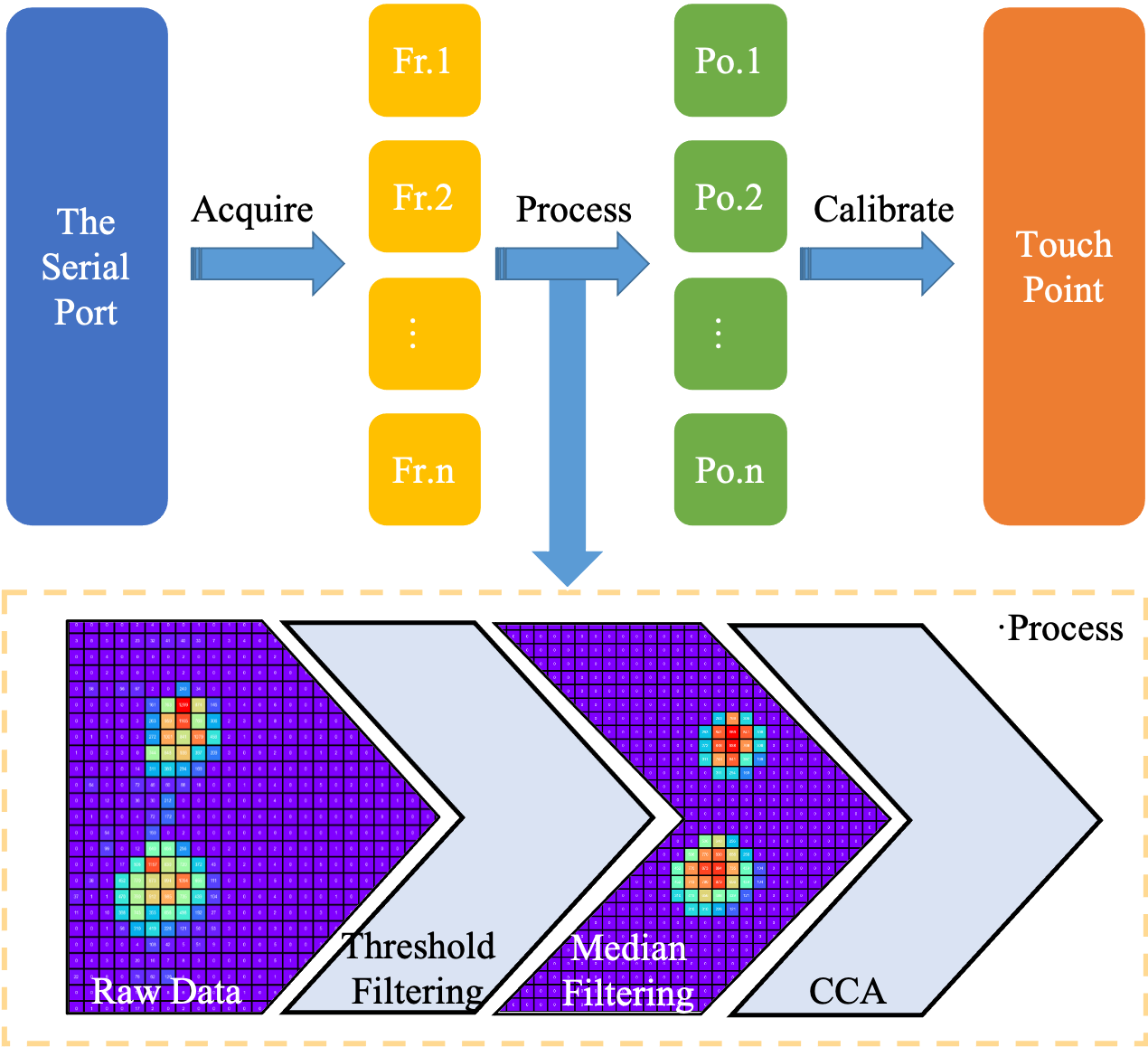}
 \caption{Illustration of touch point recognition. "Fr": Frame, "Po": Point. $n=3$ in our experiments.}
 \label{fig:gesturerecognition}
\end{figure}
Fig. \ref{fig:gesturerecognition} illustrates the gesture recognition technique used in our WristSketcher. 

\textbf{Data Acquisition and Preprocessing.} Our system reads the sensor data of each frame from the serial port of our WristSketcher in real-time, and converts the hexadecimal data into a 40 $\times$ 40 matrix. 
Since the raw data can be noisy, we also apply a threshold followed by a two-dimensional median filter \cite{ASurveyonVariousMedianFiltering} to mitigate the impact of noise. 

\textbf{Touch Point and No. of Touching Fingers.} We employ the one-scan connected-component analysis (CCA) algorithm~\cite{CCA}, which uses a randomly-selected pressed position as a seed, and then traverses its 8 neighborhoods, identifying other points pressed by the same finger if the neighbour has a pressure value greater than half of it. We repeat this process, treating each identified neighbour as a new seed until there are no new seeds, and marking all identified points as having been pressed by the same finger.
Note that we discard all identified points if their number is less than 5 to reduce the impact of noise.
For all points pressed by a single finger, we define its {\it touch point} as the one with the highest pressure value.
Then, to identify the the number of touching fingers reliably, we define it as the one with the highest frequency in consecutive frames.
We further calibrate the touch point as the average of those in these highest-frequency frames.
The final recognition results are transmitted to the AR glasses through Socket communication. 
We conducted a test to investigate whether our touch point detection method would greatly affect the performance. The test results show  that our gesture recognition runs in real-time at a rate of approximately 31 FPS.

\subsection{Gesture Recognition}
The gestures used by our WristSketcher for interaction commands can be summarized into three categories: \emph{Tap}, \emph{Double Tap}, and \emph{Long Press}. 
These three types of gestures can be distinguished according to their interval, duration, and type of leading gesture. 
Specifically, we recognize a gesture as i) \emph{Tap} if its time interval is less than 0.15s and there is not a second tap in 0.5s; ii) \emph{Double Tap} if there are two valid taps within 0.5s; iii) \emph{Long Press} if the touch time is longer than 1s.

\section{STUDY 2:USER EVALUATION}
To investigate the feasibility and usability of our WristSketcher, we conducted a user evaluation experiment on the accuracy of gesture recognition accuracy and the experience of animation creation. 
We also compared our WristSketcher with \emph{Freehand} sketching, one of the modes of mid-air bare-hand interaction, its input is made by tracking the position of the hand in real time, to demonstrate its effectiveness as a new input mode for AR glasses.

\subsection{Participants}
We recruited 10 participants (7 males, 3 females) aged 21-27 years ($\overline{x}= 23.1$, SD=1.79) for this experiment. All participants had no experience or relevant expertise in AR glasses. To reduce the effect of insufficient proficiency, the participants practiced the use of AR glasses and our WristSketcher for 20 minutes before the experiment.


\subsection{Task and Procedures}

\subsubsection{Gesture Recognition Accuracy}
We started by demonstrating to the participants the twelve different types of gestures: \emph{One-Finger Left Tap}, \emph{One-Finger Right Tap}, \emph{One-Finger Top Tap}, \emph{One-Finger Bottom Tap},  \emph{One-Finger Left Long Press}, \emph{One-Finger Right Long Press}, \emph{One-Finger Top Long Press}, \emph{One-Finger Bottom Long Press}, \emph{One-Finger Double Tap}, \emph{Two-Finger Tap}, \emph{Two-Finger Left Long Press}, and \emph{Two-Finger Right Press}. Afterwards, the participants were asked to perform each type of gesture 10 times. The order of the gesture types was randomized.

\subsubsection{Pre-defined and Freeform Creation}
Due to time constraints, we first showed the users how to use our WristSketcher for command interaction and animation, and then asked them to complete two pre-defined animation creation tasks: creating a twinkling star and a raining effect, which are the most representative and time-consuming ones in our system. 
Throughout the creative process, the facilitator recorded observations but did not intervene. After these pre-defined tasks, the participants were encouraged to use the tools freely to create. Finally, after completing all tasks, we invited the participants to rate our system on a 5-point Likert scale in terms of ease of use, usability, and functionality.

\subsubsection{Comparison with Freehand Sketching}
\begin{figure} [!h]
    \centering
    \subfloat[]{\includegraphics[width=0.45\linewidth]{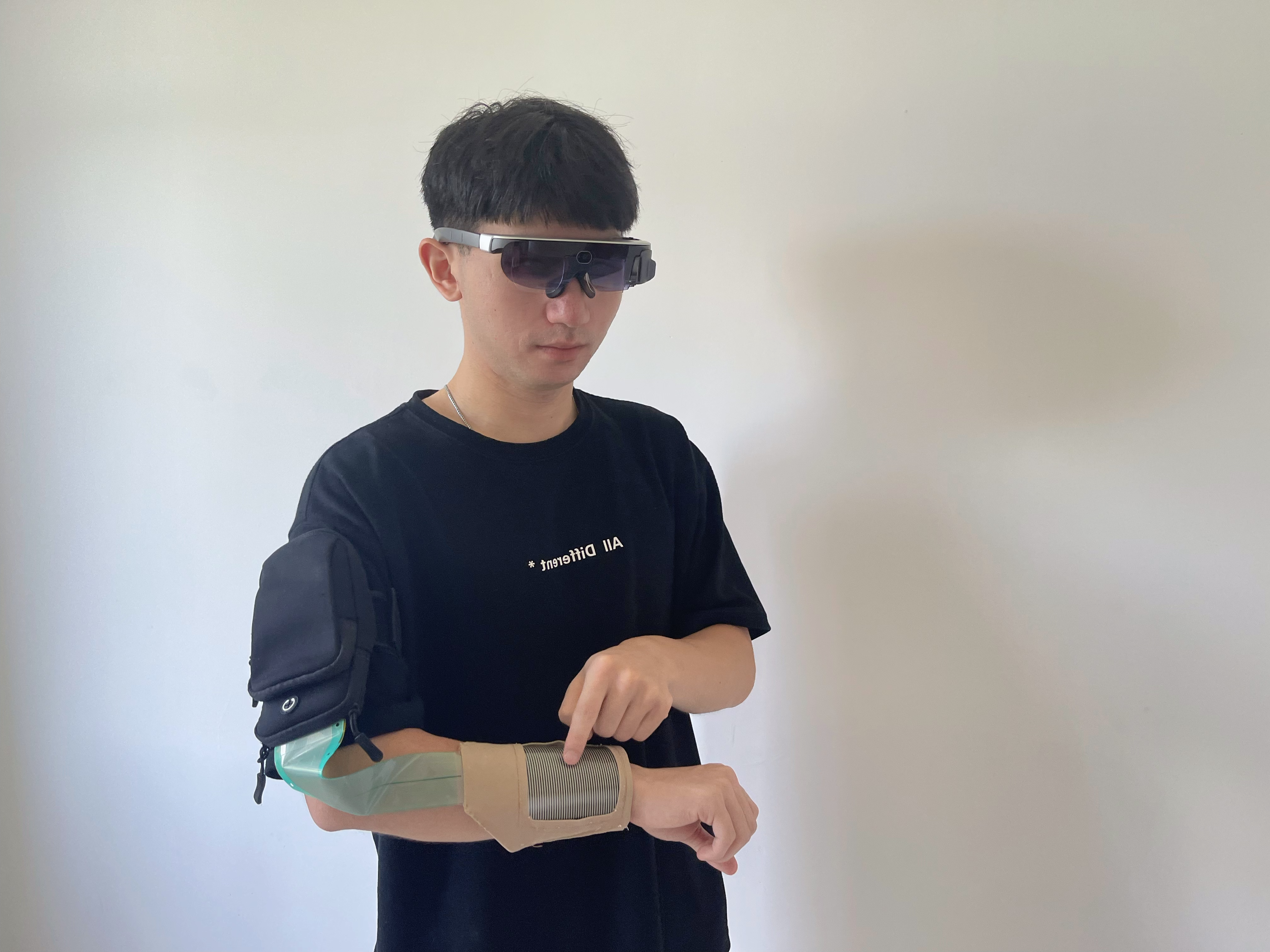}}
	\subfloat[]{\includegraphics[width=0.45\linewidth]{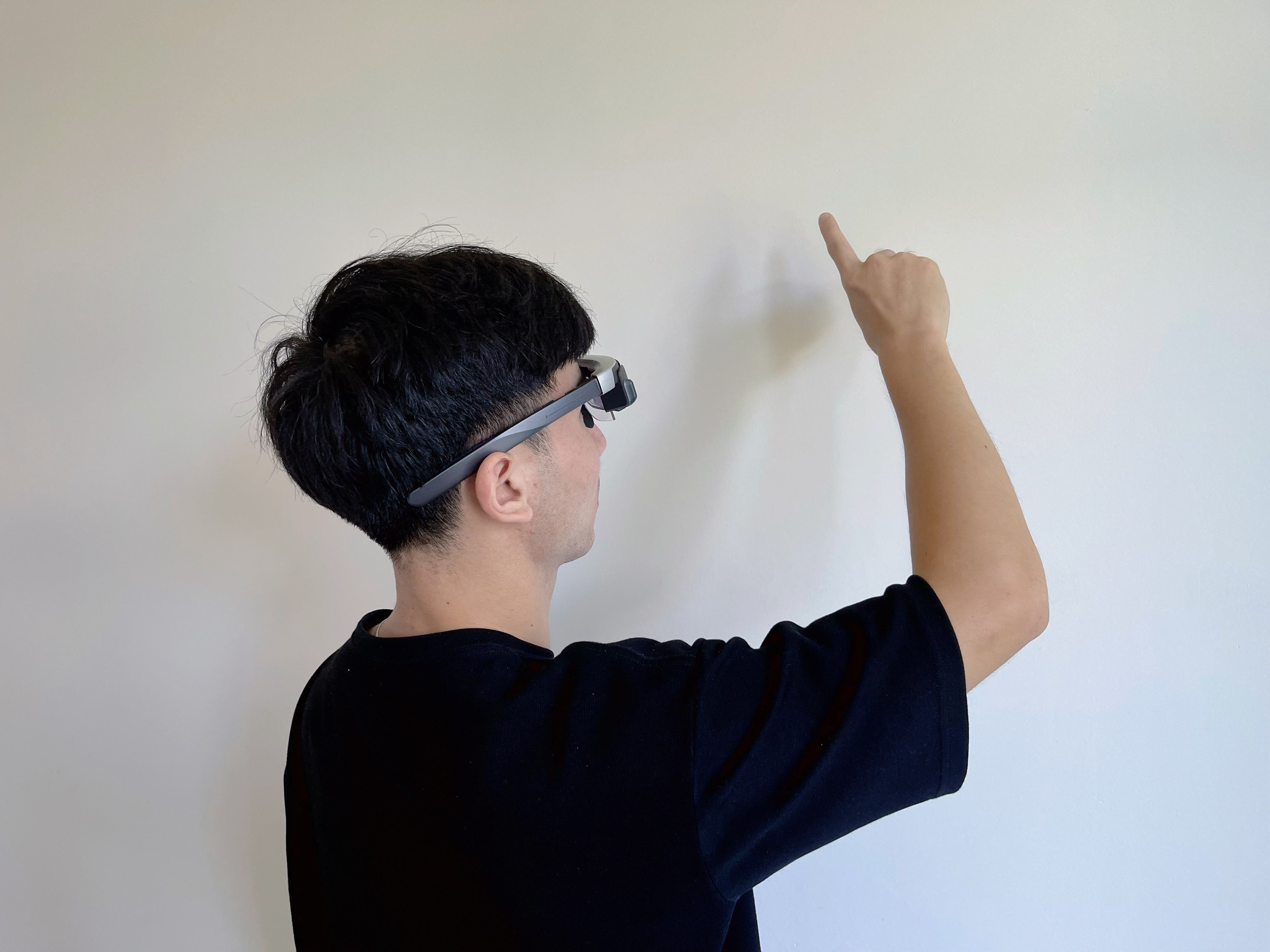}}
	 \caption{Participants drawing using (a) WristSketcher and (b) \emph{Freehand}.}
	 \label{fig:CompareFollow} 
\end{figure}

We followed the common practice in sketch interface evaluation \cite{Bare-Handed2018,VRSketchPen,InvestigatingWiese,ARSketch} and asked the participants to complete the drawing of three basic templates (Rectangle, Triangle, and Circle) in AR glasses in two input modes:
\begin{itemize}
    \item \textbf{WristSketcher.} The participants wore the WristSketcher on their arms (Fig. \ref{fig:CompareFollow}a) and input with gestures on its surface.
    \item \textbf{Freehand.} The participants used bare hands to draw continuously in the air (Fig. \ref{fig:CompareFollow}b).
\end{itemize}

During the experiment, the participants were asked to draw according to all the three templates that were presented three times each in a random order. The drawing time and results were recorded and used in the comparison of the following two metrics:
\begin{itemize}

\item \textbf{Completion Time.} This is the time spent between the first and last points of a drawing.
\item \textbf{Drawing Error.}  This is defined as the average minimum distance between the drawing and its corresponding template: 
\begin{equation}
\label{equ1}
 DE=\frac{\sum\limits_{i=1}^N ||d_i-t_i||}{N}, d_i \in D, t_i \in T, 
\end{equation}
where $N$ is the size of the sample set $D$ of the drawing and that of the sample set $T$ of $D$'s corresponding template, $t_i$ is the nearest point to $d_i$ in $T$.
\end{itemize}

\subsection{Results and Discussions}
\subsubsection{Gesture Recognition Accuracy}
\begin{figure}[tb]
 \centering 
 \includegraphics[width=\columnwidth]{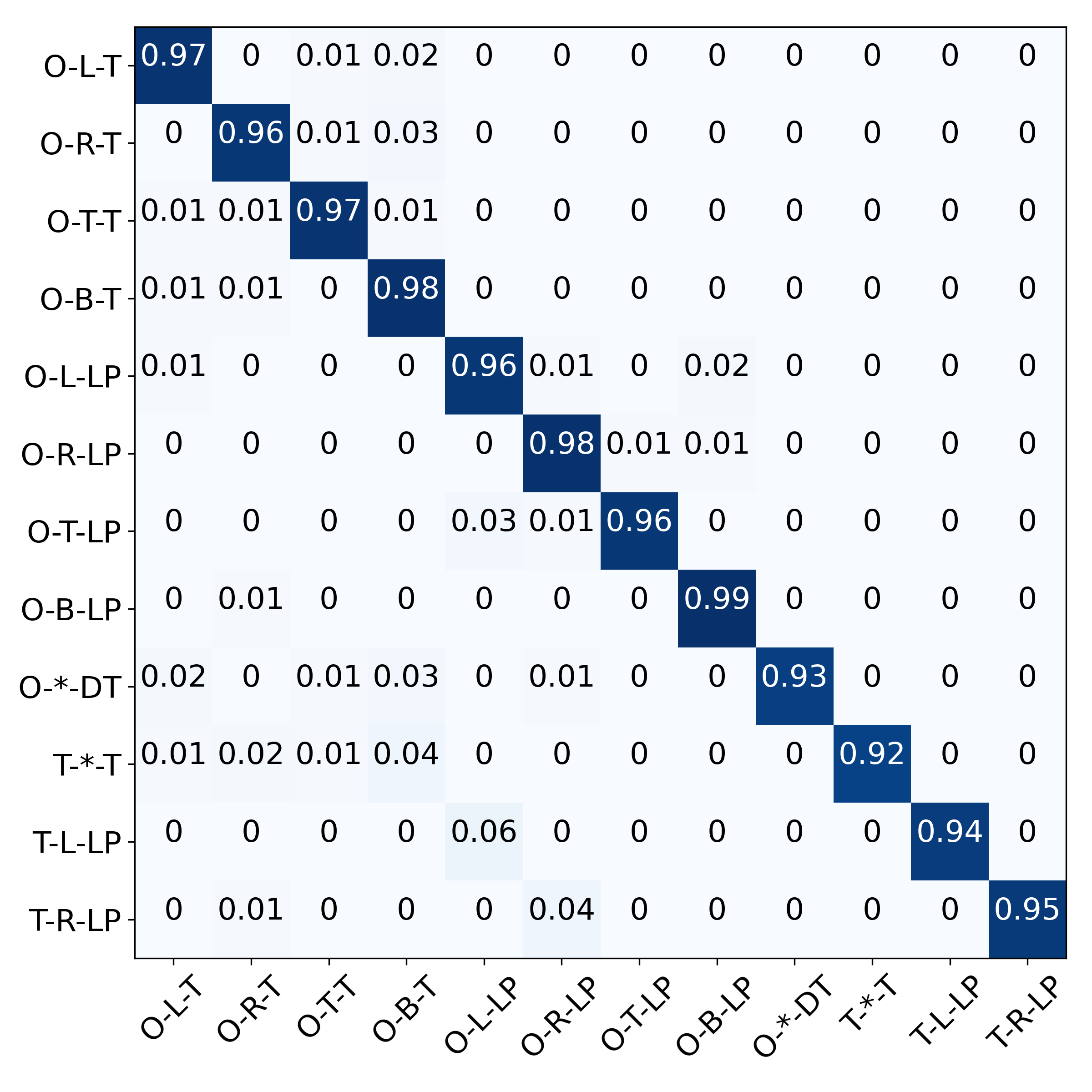}
 \caption{The accuracy confusion matrix of gesture recognition. Horizontal axis: predicted value. Vertical axis: real value. O-: One-Finger, T-: Two-Finger, -L-: Left, -R-: Right, -T-: Top, -B-: Bottom, -*-: no area limited, -T: Tap, -DT: Double Tap, -LP: -long press, e.g., O-L-T: \emph{One-Finger Left Tap}.}
 \label{fig:confusionmatrix}
\end{figure}
We define gesture recognition accuracy as the number of correctly recognized gestures over the total number of gestures.
Experimental results show that our WristSketcher achieves a high accuracy of 96.0\% (SD: 0.02).
Fig. \ref{fig:confusionmatrix} shows the confusion matrix of our gesture recognition experiment. 
It can be observed that among all gestures, \emph{One-Finger Long Press} achieves the highest recognition accuracy (97.25\%), followed by \emph{One-Finger Tap} (97.0\%) and \emph{Two-Finger Long Press} (94.50\%), while those of \emph{Two-Finger Tap} and \emph{One-Finger Double Tap} are lower (93.0\%, 92.0\%). We ascribe such lower accuracy to i) the observation that some participants occasionally touched with insufficient pressure, resulting in the sensor not being able to acquire touch information and mis-recognize the 
participants' \emph{Two-Finger} operation as a \emph{One-Finger} operation;
ii) since the recognition of \emph{One-Finger Double Tap} is based on the time interval between two successive taps and the observation that some participants occasionally failed to tap again in time, the system might mis-recognize a \emph{One-Finger Double Tap} as two \emph{One-Finger Tap} gestures.
Both cases can be mitigated by adjusting system parameters like the pressure and the time interval thresholds.

\subsubsection{Pre-defined and Freeform Creation}
Our participants were able to complete the creation of both pre-defined dynamic sketches without assistance. From the time recorded, the participants spent an average of 22.42 minutes (Min=16.93 min, Max=27.21 min) on this task, with an average of 3.77 min (Min=2.94 min, Max=4.74 min) to create a twinkling star, and with an average of 3.83 min (Min=3.14 min, Max=5.01 min) to create a raining effect. This shows that with a short practice session, the users could quickly learn to use WristSketcher to create dynamic sketches.

\begin{figure} [!h]
    \centering
	  \subfloat[]{
       \includegraphics[width=0.3\linewidth]{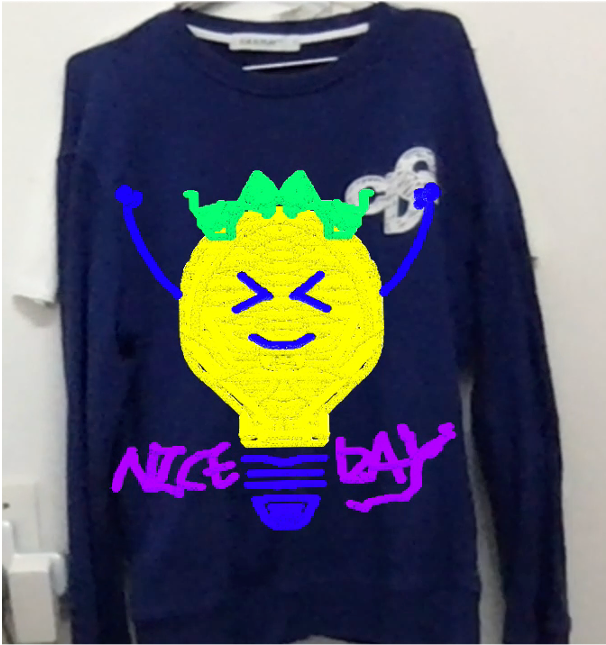}}
    \label{fig:fig11a}
	  \subfloat[]{
        \includegraphics[width=0.3\linewidth]{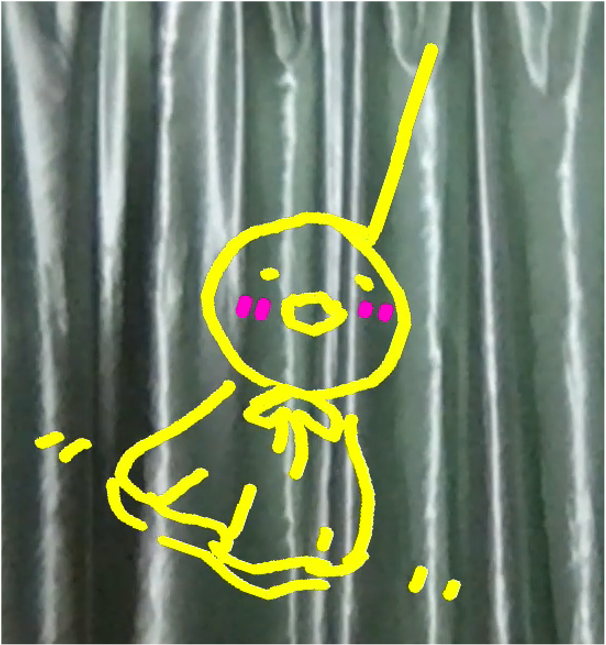}}
    \label{fig:fig11b}
	  \subfloat[]{
        \includegraphics[width=0.3\linewidth]{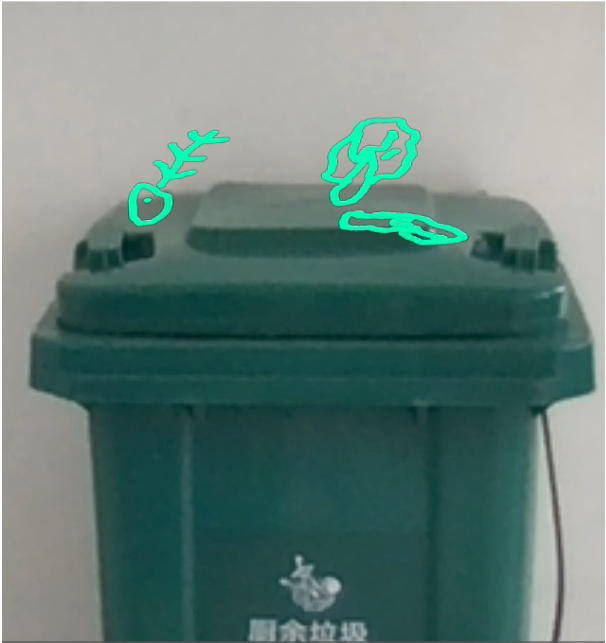}}
    \label{fig:fig11c}
	  \subfloat[]{
        \includegraphics[width=0.46\linewidth]{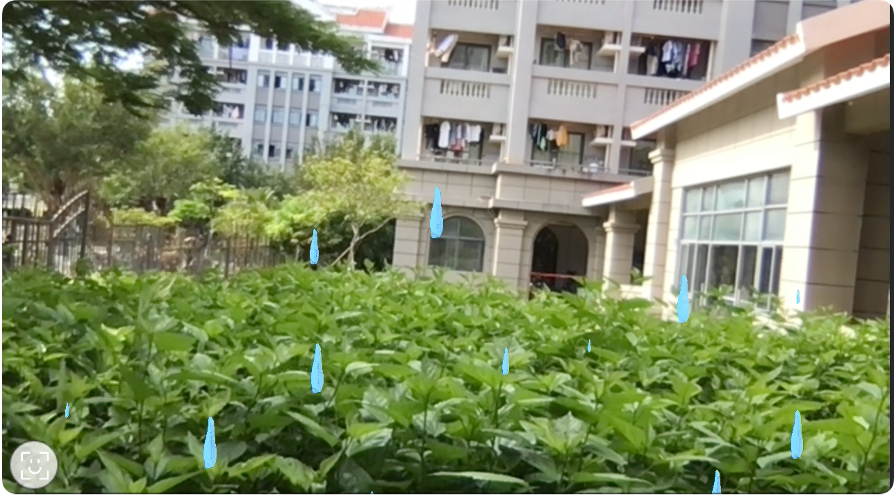}}
     \label{fig:fig11d}
     \subfloat[]{
       \includegraphics[width=0.46\linewidth]{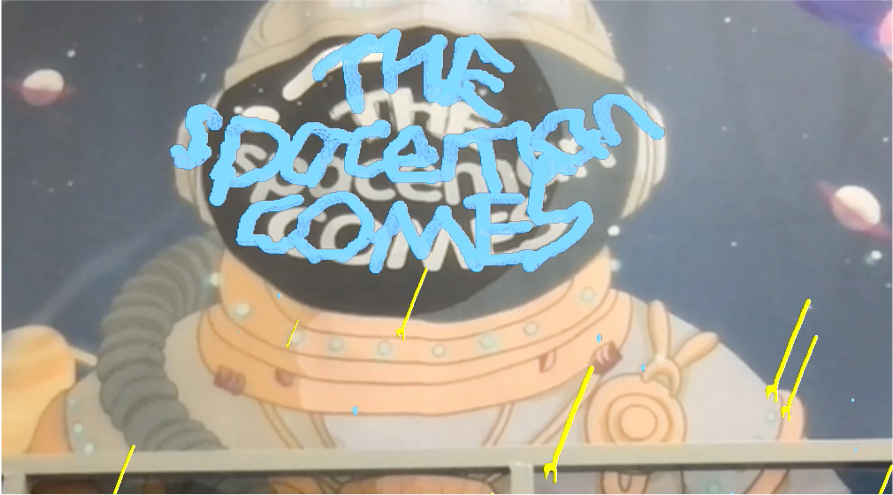}}
    \label{fig:fig11e}
	  \subfloat[]{
        \includegraphics[width=0.3\linewidth]{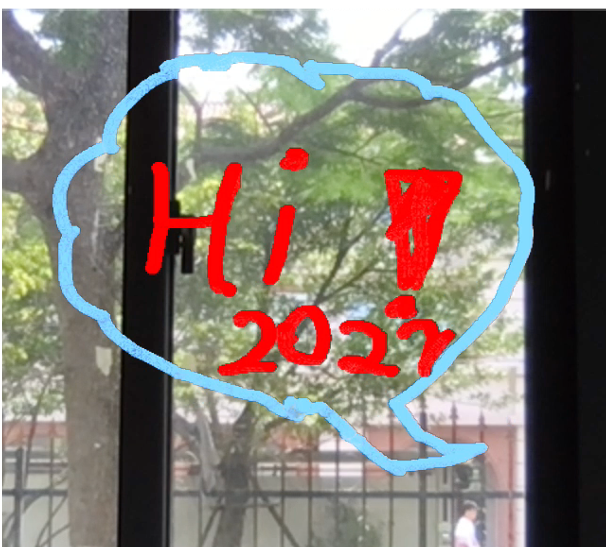}}
    \label{fig:fig11f}
	  \subfloat[]{
        \includegraphics[width=0.3\linewidth]{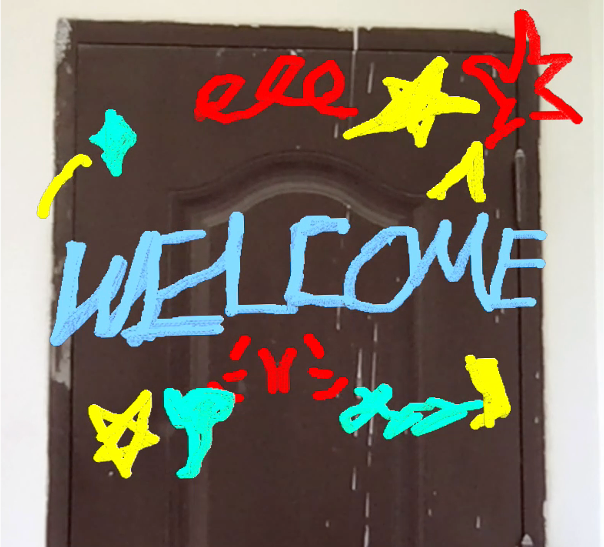}}
    \label{fig:fig11g}
	  \subfloat[]{
        \includegraphics[width=0.3\linewidth]{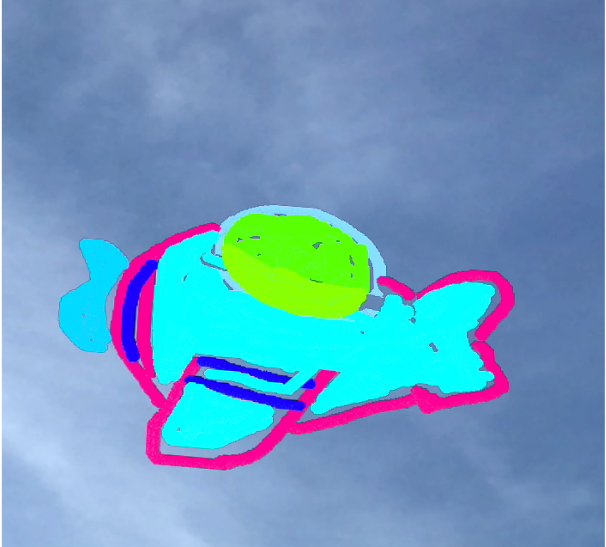}}
     \label{fig:fig11h}
	  \caption{A gallery of sketches created by our WristSketcher. The sketching times for each example are as follows: (a) 25.19 min; (b) 5.32 min; (c) 4.03min; (d) 3.12min; (e) 4.51min; (f) 5.36 min; (g) 18.06 min; (h) 20.35 min. 
	  (c),(d),and (h) were created outdoors, while the rest were created indoors.}
	  \label{fig1} 
\end{figure}

Fig. \ref{fig1} shows the freeform sketches drawn by the participants, demonstrating the innovation potentials of our WristSketcher in art creation, memory aids, and
entertainment applications. For example, Fig. \ref{fig1}a is a dynamic pattern drawn on clothes. For the intention of this design, the participant mentioned: ``My clothes are very simple and plain, and I want to design some patterns for my clothes that I like. Animate it to make it cute and fun.'' 
Fig. \ref{fig1}c is used as a memory aid for garbage types. The participant who drew it said, ``The local garbage classification is implemented, but sometimes I cannot tell which trash can I should throw it into. By doing this, when I will throw the garbage in the future, just look at the patterns I have drawn. It is a lot easier.''

At the end of the experiment, most of the participants gave positive feedback on the ease of use, usability, and functionality of our WristSketcher. From the 5-point Likert scale (1=strongly disagree 5=strongly agree), the participants' average score for the overall rating was 4.01. Specifically, the average score for ease of use was 4.10 (Min=3.5, SD=0.47), the average score for usability was 4.03 (Min=3.5, SD=0.34), and the mean score for functionality was 3.90 (Min=3.4, SD=0.30). 
In the supporting comments, the vast majority of the participants felt that the gesture design was straightforward and that it facilitated rapid interaction tasks. For example, \emph{P2} said, "I commonly use the gestures of the system, and I can quickly remember all the interaction gestures." \emph{P5} said, "Gestures are linked to the interface, and I can think of what actions I need to perform by positioning user interface elements.". 
In terms of usability, the participants expressed their acceptance and willingness to use WristSketcher for dynamic sketching in AR glasses as: i) our WristSketcher is lightweight and comfortable to wear; ii) it can give users a novel interactive experience; iii) it can respond accurately and quickly to user interactions. 
For the functionality, although some participants pointed out areas for improvement, such as \emph{P6}: "It would be nice if you could add an interactive response with me." and \emph{P8}: "The waiting time is too uncomfortable after I miss an option, I have to wait for a round to reselect.", most participants agreed that, overall, our WristSketcher provided an interesting creative support for dynamic sketching using AR glasses.

\subsubsection{Comparison with Freehand Sketching}

Tables \ref{table:Followtemplate1} and  \ref{table:Followtemplate2} show the qualitative and quantitative comparisons between our WristSketcher and \emph{Freehand} respectively. 
The significance of quantitative results was analyzed by one-way ANOVA. 
It can be observed that: i) for completion time, the difference between WristSketcher and \emph{Freehand} is significant (F=13.39, P$<$0.05) that sketching with our WristSketcher consumes slightly more time due to its smaller interaction area;
ii) for drawing error, the difference between the two input methods is more significant (F=31.45, P$<$0.001) that our WristSketcher allows for more accurate sketching. We ascribe this to the observation that although the user can trace standard elements steadily at the beginning, their hands become unstable over time.
Such a trade-off shows that our WristSketcher can be used as a new input mode for sketching with AR glasses.
\begin{table}[!h]   
\centering
\caption{
Qualitative comparison between our WristSketcher and \emph{Freehand}. The images show the overlay sketches by all users.
}  
\begin{tabular}{|c| c|c|c|} 
\hline  Method & Rectangle & Triangle & Circle \\   
\hline  \emph{WristSketcher} & \begin{minipage}[c]{0.2\columnwidth}
		{\includegraphics[width=\linewidth]{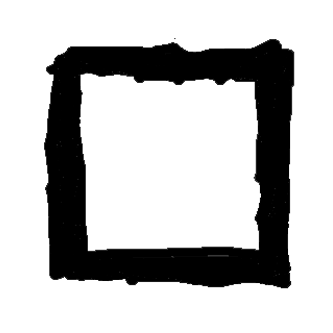}}
	\end{minipage} &\begin{minipage}[c]{0.2\columnwidth}
		{\includegraphics[width=\linewidth]{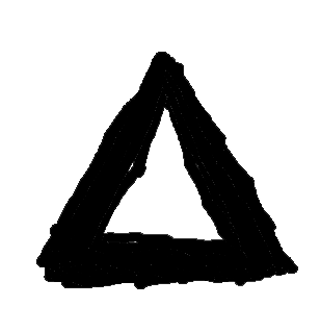}}
	\end{minipage} &\begin{minipage}[c]{0.2\columnwidth}
		{\includegraphics[width=\linewidth]{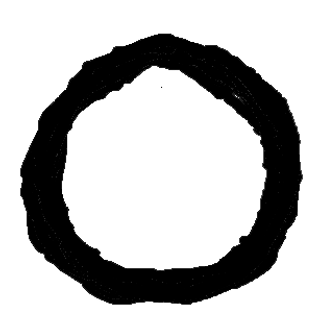}}
	\end{minipage}\\
\hline  \emph{Freehand} &  \begin{minipage}[c]{0.2\columnwidth}
		{\includegraphics[width=\linewidth]{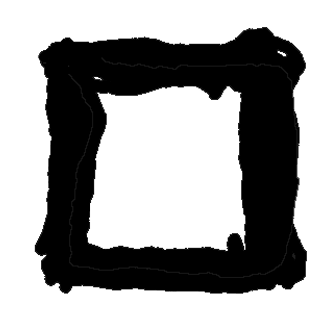}}
	\end{minipage}& \begin{minipage}[c]{0.2\columnwidth}
		{\includegraphics[width=\linewidth]{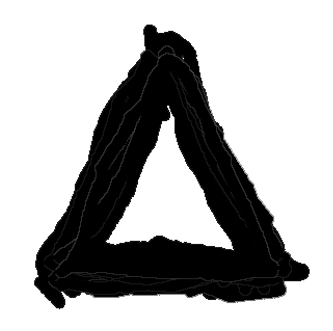}}
	\end{minipage}&  \begin{minipage}[c]{0.2\columnwidth}
	    {\includegraphics[width=\linewidth]{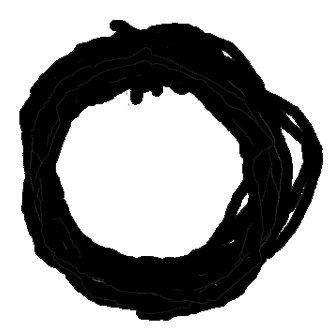}}
	\end{minipage} \\
\hline
\end{tabular}
\label{table:Followtemplate1} 
\end{table}

\begin{table}[!h]
\centering
\caption{Quantitative comparison between our WristSketcher and \emph{Freehand}.}  
\begin{tabular}{|c| c|c|} 
\hline  Condition & Completion Time (s) & Drawing Error (pixel) \\   
\hline  \emph{WristSketcher} & 13.38 $\pm$ 1.77 & 5.14 $\pm$ 2.15\\
\hline  \emph{Freehand} & 10.31 $\pm$ 1.98&  7.16 $\pm$ 2.65\\
\hline
\end{tabular}
\label{table:Followtemplate2}
\end{table}



\section{CONCLUSION AND FUTURE WORK}
We have presented WristSketcher, a new input agent that allows users to create 2D dynamic sketches in AR glasses. 
Specifically, our WristSketcher makes up for the lack of precision caused by mid-air bare-hand sketching, as well as the portability and comfort that users need to consider when using mobile devices, by diverting AR sketch input to a lightweight and flexible sensing wristband.
To implement such an input mode, we designed a set of interaction gestures through a heuristic study and its corresponding gesture recognition method.
We also extends our WristSketcher with the ability of animation creation, allowing it to create dynamic sketches embedded in real-world scenes.
Experimental results show that our WristSketcher i) achieves a high gesture recognition accuracy of 96.0\%; ii) achieves higher accuracy compared to \emph{Freehand} sketching; iii) achieves high user satisfaction in ease of use, usability and functionality; and iv) shows the innovation potentials of our WristSketcher in art creation, memory aids, and entertainment applications.

{\bf Limitations and Future Work.} 
Although our WristSketcher allows users to create dynamic sketches in  real-world scenes, the animations are not responsive to user interactions. 
We hope to extend our WristSketcher and incorporate multimodal interaction technologies for responsive animations in our future work. 
In addition, due to the hardware of AR glasses, we cannot obtain depth information and cannot use the Self-Assisted Localization (SLAM) function. While the Vuforia-based rendering technique enables the positioning of sketches in the real world, it is based on feature point matching in a view and does not allow for multiple views of the sketch drawn. So the sketches currently created by our WristSketcher are essentially 2D images. We hope to extend our WirstSketcher for 3D dynamic sketching in future work.






%

\bibliography{reference}{}
\bibliographystyle{IEEEtran}

%








\end{document}